\documentstyle[12pt]{article}
\input epsf
\def\beq{\begin{eqnarray}}   \def\eeq{\end{eqnarray}}
\def\s{\sigma_{\rm DW}}

\def\lsim{\mathrel{\rlap{\lower3pt\hbox{\hskip0pt$\sim$}}
    \raise1pt\hbox{$<$}}}         
\def\gsim{\mathrel{\rlap{\lower4pt\hbox{\hskip1pt$\sim$}}
    \raise1pt\hbox{$>$}}}         

\begin{document}
\begin{flushright}
NYU-TH/01/01/05  \\
TPI-MINN-01-03\\
UMN-TH-1935
\end{flushright}

\vspace{0.1in}
\begin{center}
\bigskip\bigskip
{\large \bf Ultralight Scalars and Spiral Galaxies}

\vspace{0.5in}      

{Gia Dvali$^1$, Gregory Gabadadze$^2$, and M. Shifman$^2$ }
\vspace{0.1in}

{\baselineskip=14pt \it 
$^1$Department of Physics, New York University,  
New York, NY 10003 \\
$^2$ Theoretical Physics Institute, University of Minnesota, 
Minneapolis, MN 55455}  
\vspace{0.2in}
\end{center}

\vspace{0.9cm}
\begin{center}
{\bf Abstract}
\end{center} 
\vspace{0.1in}

We study some possible astrophysical implications of a very weakly 
coupled ultralight dilaton-type scalar field. Such a field may develop
an (approximately stable) network of domain walls. The domain wall
thickness is assumed to be comparable with the thickness of the 
luminous part of the  spiral galaxies. The walls provide trapping for 
galactic matter. This is used to motivate the very existence of the spiral
galaxies.  A zero mode existing on the  domain wall is a massless scalar 
particle confined to $(1+2)$ dimensions.  At   distances much larger than 
the galaxy/wall thickness,   the zero-mode exchange generates a 
logarithmic potential, acting as an additional term with respect to 
Newton's
gravity. The logarithmic term naturally leads to constant rotational
velocities at the periphery. We estimate the scalar field coupling to the
matter energy-momentum  tensor  needed to fit the observable flat
rotational curves of the  spiral galaxies. The value of this coupling 
turns out to be reasonable -- we find no contradiction  with the existing
data.

\newpage

\section{Introduction}

Many years ago, Zeldovich, Kobzarev, and Okun considered \cite{ZKO}
a network of domain walls in the Universe, which inevitably
develops provided the underlying theory has a spontaneously
broken discrete symmetry, and came to the conclusion that
such a network is ruled out because its existence would lead to an
unacceptable cosmology (for a review and more detailed discussion see 
\cite{Vilenkin}).  In this paper we revisit the issue of the domain
wall network in a specific context of an ultralight dilaton field with an
ultraweak self-interaction and a universal coupling to the matter fields. We
show that such domain walls are not only compatible with experiment, but
they may be used to explain  salient features of the observed world --
the flatness of the spiral galaxies and the constancy of the rotational
velocities at the periphery of the spiral galaxies. Our suggestion can be
summarized as follows. The mechanism to be discussed below
  provides a light trapping for the galactic matter.
One or several spiral galaxies may be embedded in one wall.
The transverse width of the luminous disk of the galaxy
(which includes the major part of dark baryon matter) is of
the order  of the width of the domain wall, which is, in turn,
of the order of the inverse mass of the dilaton,  
$\Delta \sim m_d^{-1} \sim10^{20}$ cm.   At distances much larger 
than the disk thickness, 
$L\gg \Delta$, the domain wall can be approximated
by a two-dimensional surface. There is a scalar field which is 
confined to and   propagates along  this surface. The existence of this 
2+1 dimensional massless scalar is an unavoidable consequence of the
presence  of the wall itself (the zero mode).  The zero mode exchange at
distances  $L\gg \Delta$  gives rise to a logarithmic  modification of the
gravitational potential, so that the rotational velocities are constant at 
distances of a few dozen kiloparcecs, as required by  observations. 

Our model includes the usual fine-tuning of the
cosmological constant. We have nothing to add in this respect, and assume
that the cosmological constant is somehow fine-tuned to fit the
current estimate,
\beq
\varepsilon \lsim (10^{-3} \,\,{\rm eV})^4\,.
\label{two}
\eeq
The extremely small dilaton mass and couplings are then automatically 
protected by Eq. (\ref{two})
against renormalizations due to virtual matter fields propagating in
loops. Quantum gravity loops may pose a problem, see Sec. 2.

\section{Ultralight Scalars}

\vspace{0.2in}

Fundamental particle physics theories often deal with  some very weakly
coupled massless scalar particles,  such as   dilatons or other moduli. Usually
there are dynamical reasons which protect the masslessness of these fields. 
As long as they stay  massless  they could mediate gravity-competing
forces. This requires   the strength of their interactions to be adequately
suppressed.

Although the massless limit is a useful theoretical laboratory,
in actuality, the dilaton-type fields cannot remain massless --
for stabilization a mass term has to be
 generated. Usually, the mass generation
(moduli stabilization)  is attributed to some nonperturbative
physics responsible for supersymmetry (SUSY) breaking. In view
of quantum corrections, it is naturally to expect then the  mass  in 
the ballpark 
$m \sim M_{\rm SUSY}^2/M_P$, where $M_{\rm SUSY}$ is the scale
of supersymmetry breaking and $M_P$ is the Planck mass.
Since, $M_{\rm SUSY}\gsim 1$ TeV, 
the mass  comes out $\gsim 10^{-3}$ eV. 

For the applications to be discussed below we will need a much lighter 
scalar field. {\em A priori}
it is not difficult to imagine the existence of a field
with  the bare mass
\beq
m_0~\sim M_P ~{\rm exp} \left (- {8\pi^2\over g^2}  \right )~,
\label{exp}
\eeq
where $g$  is some  coupling constant, $g\lsim 1$. If, for instance,  
$g\simeq 0.8$, then we get $m_0\sim 10^{-33}$ GeV (our favorite number, 
see  below). The problem is that the bare mass that small
will not be protected from huge renormalizations in loops.
 In the full interacting  quantum theory, if there is no special
reason, the scalar mass   will be   
  shifted   towards the natural estimate
$\sim M_{\rm SUSY}^2/M_P$ by loop corrections. The question is
whether there exists a set-up that would naturally protect
a scalar field from  renormalizations.

One can pose this question at two different levels. First,
one can switch off gravity and consider loops generated by virtual particles
of the Standard Model or other fields that may be present in 
extensions of the Standard Model (these quantum corrections
will be referred to as SM loops). We  found
a mechanism protecting against these SM loops.
A model will be presented below
 in which the mass and coupling constants
of a universally coupled scalar field
are  protected in a natural way.

At the next level, one can start bothering about gravity-generated loops.
Generally speaking, the
mass of any elementary scalar field is not  protected
from renormalizations by quantum gravity loops. Thus, a protection
against the quantum gravity loops may be needed -- perhaps,
a fine-tuning of a special kind. Since quantum gravity is not yet a closed
theory, we will not ask  in the present
paper what is the precise nature  of this protection, but, rather, focus on
 astrophysical consequences of an ultralight  
universally coupled scalar field.

(Let us parenthetically note that an  obvious possibility is to assume
that the scalar field in  question is a composite one,  with a very low
compositness scale.)

The  model we suggest is as follows.
 Consider a field theory which has
both dimensionless and dimensionful parameters (the latter include,
in particular, the physical masses as well as the ultraviolet regulator
masses). In spite of the presence of the dimensionful parameters, one can
make the theory scale-invariant if every parameter of mass dimension
$\nu$ is multiplied by 
\beq
\exp \left(\frac{\phi}{M_*} \nu\right)\,,
\eeq
where $\phi$ is a dilaton field.
 The scale transformation
\beq
x\to \kappa\cdot  x\,,\quad \Phi\to \Phi \cdot (\kappa)^{-{\rm
dim}_\Phi}\,,
\eeq
where $\Phi $ is a generic field and ${\rm dim}_\Phi$ is its dimension,
must be  supplemented by a shift of the dilaton field,
\beq
\phi \to \phi + M_* \ln \kappa\,.
\eeq
The scale invariance is then obvious.

As long as the scale invariance is unbroken,
the effective Lagrangian of the dilaton field, all loops included,
must be proportional to 
\beq 
\varepsilon \, \exp \left(\frac{4 \phi}{M_*} \right)\,.
\eeq
If the vacuum energy density $\varepsilon$ is fine-tuned as in Eq.
(\ref{two}) and $M_*>M_P$ (as will be the case, see below),
the loop renormalizations of the quadratic, cubic, quartic, etc. terms in the
dilaton field will be negligibly small, automatically.
When we give a mass to the dilaton field, and some self-interactions, we
explicitly break the scale invariance in the dilaton sector.
If the coupling constant is sufficiently small, however,
this breaking will not lead to large renormalizations either.
A typical constraint is
\beq
\lambda \, M_{\rm SUSY}^2 < m_d^2\,,
\label{three}
\eeq
where $\lambda$ is a quartic coupling constant. 
Equation (\ref{three}) implies that $\lambda \ll 10^{-72}$.
In fact, in our model  $\lambda$ is many orders of magnitude
smaller than the estimate above. 

The coupling of the dilaton field under consideration to matter is 
universal, through  the
trace of the matter  energy-momentum tensor  $\theta_{\mu\nu}$,
\beq
\Delta {\cal L}_{\rm int} =  {\phi \over M_*}~\theta^\mu_\mu\,.
\label{one}
\eeq 
We will refer to any such  $\phi$'s as the dilatons.\footnote{Note that the
string theory  dilaton does not have such a property and, unless it is heavy, 
 gives rise to the violation 
of the equivalence principle even in the nonrelativistic approximation
\cite {Veneziano}.  We will  not  deal with the string dilaton, but, rather, 
concentrate on a hypothetical dilaton
 which couples universally to the trace 
of the matter energy-momentum tensor. This latter can lead to the 
equivalence-principle violating effects in relativistic experiments
(see below).}. 
In this language, say, the loop correction to the dilaton mass
 is determined  by the  zero-momentum 
two-point function of $\theta^\mu_\mu$'s
\beq
(M_*)^{-2}\, \int ~d^4x ~\langle \theta^\mu_\mu (x) 
~\theta^\nu_\nu(0)~\rangle~
=(M_*)^{-2} (-4) \,\langle \theta^\mu_\mu \rangle~.
\label{corr}
\eeq 
The proportionality of this correlator to the vacuum expectation 
value of $\theta^\mu_\mu$ is a consequence of the 
scale Ward identity 
\cite {SVZ}.
Therefore, the mass renormalization  is 
proportional to the cosmological constant,
\beq
\Delta\, ( m_d^2) \sim \varepsilon/M_*^2\,,
\eeq
 and can  be neglected provided the cosmological constant is fine-tuned to
its empiric value.

In this way,  
we naturally arrive at an ultralight scalar;\footnote{Mind the issue of
gravity loops, see above.} its
 Compton wavelength may be of the astronomical (galactic) size
$\sim 10^{20}$ cm (corresponding to the mass
$m_d\sim 10^{-33}$ GeV).  The parameter $M_*$ will be fitted below
to generate the observable value of the tails of the rotational curves
in the spiral galaxies,
\beq
M_* \sim \sqrt{\frac{M_G\, m_d}{C}}\,,
\label{five}
\eeq
where $M_G$ is the mass of the baryon matter in galaxy,
 $M_G\sim 10^{68} $ GeV, and $C$ is a dimensionless constant of
the  order of
several units times $10^{-7}$ (see Sec. 4).
Numerically, Eq. (\ref{five}) implies that
$M_*$ is 10 to 100 Planck masses. It is
remarkable that a constant of  the Planck scale emerges from
such macroscopic quantities as the galactic mass and its width
$\Delta\sim m_d^{-1}\sim 10^{20}$ cm.

At distances much less than $\Delta$ the dilaton can be considered as
massless; it will provide a long-range force competing with the Newton
gravity, with a different vertex structure (the scalar rather than tensor
exchange).
 Needless to say that this   violates the equivalence
principle in relativistic gravitational measurements. Therefore, 
its coupling must be  sufficiently suppressed.
We will discuss this issue at length below. Here we only note that
if the parameter
$M_*$ is larger than
\beq
M_*^2 \gsim (10^{3} - 10^{4})~M_P^2,
\label{bound}
\eeq 
the effects of the equivalence principle violation
are below the currently observable level. 
Remarkably, the very same value of $M_*$ emerges from the
domain wall explanation of the constant tails of the
rotational curves.

In the present paper we will adopt a pragmatic
approach  and study  astrophysical
implications of such very weakly interacting and
extremely light dilaton  field(s) universally
coupled to the trace of the energy-momentum
tensor, without
speculating on its possible origins or trying to explain, from   first
principles,
why its mass might lie in the ballpark of $10^{-33}$ GeV.
We will postulate a specific  shallow self-interaction potential.  It must
exhibit a discrete symmetry, which will be spontaneously broken in the
vacuum state. Then a network of domain walls can develop. As we will see,
this may lead to trapping of the galactic matter  within the wall
world-volume and, in addition, to a logarithmic  long-range interaction
potential on the surface of the domain wall. Our goal is 
to explore whether
such  set-up can provide an explanation for the existence of 
the spiral 
galaxies  and for the constancy of the rotational curves of the 
spiral galaxies. Before passing to  astrophysical aspects,
we briefly review domain walls in field theory.
Our purpose is to formulate what is needed from the dilaton
sector to produce the domain walls of the required type.

\section{Domain Walls}

To begin with, let us assume that we deal with one real field
$\phi$, and switch off  the gravitational coupling.
The above simplifying assumptions allow us to present  the main idea
without inessential complications.  
After the mechanism  is presented,  we will consider a two-field model
and will switch on  the coupling  with  gravity.

Let us consider the simplest possible Lagrangian leading to domain walls
(for a review of domain walls in the given context, see e.g. Ref.  \cite
{Cvetic}),  
\beq
{\cal L}~=~{1\over 2} \left ( \partial_\mu ~\phi \right )^2~-~
{\lambda\over 2}~\left (\phi^2 ~-~\eta^2   \right )^2\, .
\label{lagr}
\eeq
As we will see below, this Lagrangian is not
quite realistic; a
more involved model is needed to ensure both, the matter trapping and
the proper logarithmic potential.
However, the simple Lagrangian below   demonstrates
the main idea in terms of analytic expressions (while in more
realistic cases  only numerical solutions can be given). 

We temporarily drop interactions of $\phi$ with the matter fields.
The classical equations of motion of this system possess a  solution
in the form of a domain wall  stretched in the 
$x,y$ plane,
\beq
\phi_{\rm cl}(z)~=~\eta~{\rm tanh}~\left (\sqrt{\lambda}~
\eta~z \right ) .
\label{cl}
\eeq
The $\phi$ mass $m$ and the wall tension $\s$ are
\beq
m=2\sqrt{\lambda}\eta\,,\qquad \s =4\sqrt{\lambda}\eta^3/3\,.
\eeq

Let us consider perturbations around this solution.
It is well known that there exists a massless 
mode $\rho$ which lives on the wall
worldvolume. This can be expressed as follows:
\beq
\phi(t,x,y,z)~=~\phi_{\rm cl}(z)~+~{1\over \sqrt{\s}}~
{d\phi_{\rm cl}(z)\over dz}~\rho(t,x,y) .
\label{phi}
\eeq
One can check that
$\rho$ satisfies the  equation of motion for a free massless
(2+1)-dimensional scalar field,
\beq
\left (\partial_t^2~-~\partial_x^2~ -
~\partial_y^2 \right )~\rho(t,x,y)~=0\,.
\eeq
Therefore, the field $\rho$, when coupled to matter appropriately,
  mediates logarithmic 
potential between sources located on the 
domain wall,
\beq
\langle~ \rho(0,x,y)~\rho(0,0,0) ~\rangle 
~\sim ~{\rm ln}\sqrt{x^2~+~y^2}~.
\eeq
 The logarithmic potential at  large distances
takes over the Newtonian $1/r$, with necessity,
and generates a constant  component in the average rotational
velocity (i.e. a component which does not fall off with the distance $r$). 
From this point of view, the domain wall built from one real
scalar field $\phi$ is perfectly sufficient.
What makes us consider more complicated dilaton sectors?

There are two reasons. First, we would like the walls to act
as a natural  trap  for matter. Given the profile of the field
$\phi$ in the wall of the type (\ref{cl}) we see that the coupling (\ref{one})
leads to a monotonous variation of the
nucleon masses -- the nucleons are slightly heavier
on one side of the wall and slightly lighter on the other.
This is not what we want. We want them to be lighter inside the wall,
and heavier outside. To this end the profile of the
field coupled to $\theta_\mu^\mu$ must be of the type depicted on Fig. 1.

\begin{figure}   
\epsfysize=7cm
\centerline{\epsfbox{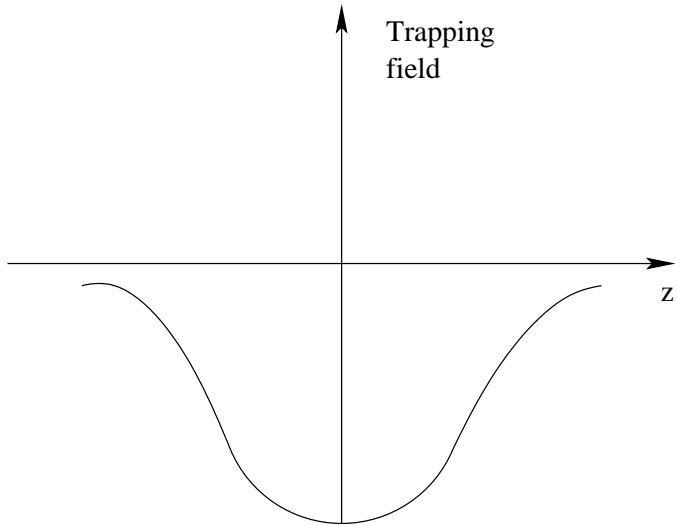}}
 \caption{The bell-shaped profile of the field modulus.}
\label{Fig.2}
\end{figure} 

It is not difficult to achieve such a profile in two-component models,
where, in fact, it is quite typical. Assume we have two real fields,
$\phi$ and $\chi$, the ``primary" domain wall of the field $\phi$ is of the
type (\ref{cl}), and it forces a structure in the $\chi$ field of the type
\beq
\chi_{\rm cl}(z)~\propto -\frac{1}{{\rm cosh} \left (m z \right )}
\eeq
to develop. This is the case, e.g., in the potentials of the type
\beq
V(\phi, \chi ) =\left(\frac{m^2}{\sqrt{\lambda}} - \sqrt{\lambda}\phi^2
-\alpha\chi^2
\right)^2 +\beta\phi^2\chi^2\,,\quad \alpha ,\beta > 0\,,
\eeq
which were recently considered
\cite{SSV}  in a different context. The maximal $|\chi |$ is achieved at the
point  right in the middle of the $\phi$ wall, where $\phi = 0$.
Denote this maximal value by $\eta_\chi$. For our purposes
it is necessary that $\eta \ll \eta_\chi$ while $m_\chi \ll m_\phi$
where much less can actually mean a factor of several units.

Assume that both, $\chi$ and $\phi$ are universally coupled to the trace of
the energy-momentum tensor,
\beq
\Delta {\cal L}_{\rm int} =  \left( {\phi \over M_*}
+  {\chi \over M_*}\right)\theta^\mu_\mu\,,
\label{six}
\eeq 
and the constraints above are satisfied.

The main virtue of this set-up is the possibility 
to trap the galaxy matter inside the domain wall. 
Since $\eta \ll \eta_\chi$ the main role in the trap
is played by the $\chi$ field.
  The  matter  interaction  (\ref{six})
is such that nucleons are lighter inside the wall.
This can be viewed as a shallow potential \cite {Volosh}
for particles which constitute the galaxy. 
The galaxy matter is naturally  lightly 
trapped. The relative nucleon mass variation inside/outside
can be easily estimated to be $\delta M/M \sim 
10^{-9}$ to $10^{-10}$. The corresponding
escape velocity in the direction perpendicular to the wall (galaxy) plane
is  3 to 10 km/sec.

At the same time, since  $m_\chi \ll m_\phi$
the zero-mode mediating logarithmic potential will be predominantly
associated with the field $\phi$, rather than $\chi$.
We need this to be the case since it is $d\phi /d z$
that has the proper bell-like shape, rather than  $d\chi /d z$.
The  bell-like shape of the zero mode is needed in order to ensure
such (logarithmic) attraction of the distant bodies in the galaxy lying
inside the domain wall which would be (approximately)
independent of the position of the body in the transverse direction.
The profile of  $d\chi /d z$ is inappropriate for that purpose.

Another reason for playing with more sophisticated
dilaton sectors is the desire to have a network of domain walls
with the stable wall junctions. The stability of the wall junctions cannot be
attained in the simplest ${\bf Z}_2$-based models.

In order to produce the junction solutions one should start from 
a more complicated  Lagrangian for a scalar field, with
some ${\bf Z}_N$ symmetry.
A simplest possibility 
along these lines would be 
to start from  a ${\bf Z}_4$ symmetric potential for a complex 
scalar field $\varphi$,
\beq
V(\varphi^+,\varphi)~=~{\lambda\over M_p^4}\left |\eta^4~-~
\varphi^4 \right |^2~.
\label{z4}
\eeq
There are four minima described by this potential:
$\varphi_{\rm vac} = \eta ~{\rm exp}\left (i2\pi k/4 \right ),
~~k=0,1,2,3.$
The original ${\bf Z}_4$ symmetry is spontaneously broken
in any of these four vacua. As a result, ${\bf Z}_4$ 
domain walls  and junctions could  form (see Fig. 2).
Each wall in the junction  can move along the transverse direction.
Hence, there are zero modes on each of them. These latter 
remain to be zero modes
as long as gravitational interactions are neglected. (The corresponding
mixing
of the zero modes with gravitons is discussed below; it turns out to be
unimportant.) 

\begin{figure}   
\epsfysize=7cm
\centerline{\epsfbox{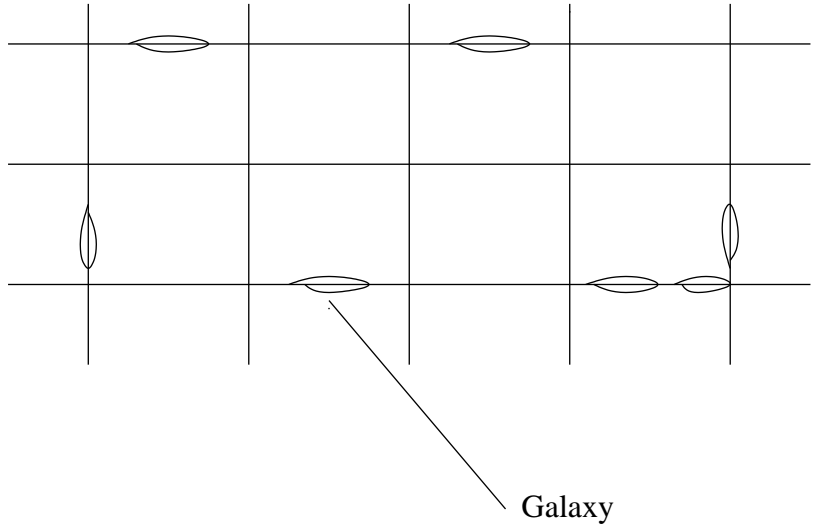}}
\caption{A network of domain walls with immersed galaxies.}
\label{Fig.1}
\end{figure} 

The idea is that the
spiral galaxies will be located on the links between the junctions.
On the other hand, the junction points can be used
to create centers for elliptic galaxies. Of course, we imply that in the
realistic situation the domain wall network is more like a soap foam 
than the regular lattice of Fig. 2.

So far, in discussing the wall structure we ignored the facts that
the dilaton fields are coupled to gravity and to matter,
i.e. we ignored the
back reaction of matter on the walls themselves.

What changes when the
interaction of the scalar field with gravity is switched on?
Had we an idealized case of a single wall 
with a single galaxy on it, then the regions of the wall 
where there is no galaxy matter 
(i.e., far away from the galaxy disc along the wall)
would inflate with the following line element 
\cite {Vilenkin,IpserSikivie}:
\beq
ds^2~=~\left (1-H~|z|\right )^2~dt^2~-~dz^2~-~
\left (1-H~|z|  \right )^2~e^{2Ht}~\left (dx^2~+~dy^2\right )~.
\label{inflation}
\eeq
The inflation  would be governed by a Hubble parameter which is 
defined as $H\sim G_N \s$. Below we will argue that
the wall tension $\s$ must  be smaller than $$\s \ll 
10^{-5}~{\rm GeV}^3\,, $$
 so that  $H\ll
10^{-43}~{\rm GeV}$. Therefore, the horizon in the $z$ direction,
which is due to the Rindler type metric of the domain wall, 
would be well beyond our true horizon.
In reality, however, there are other domain walls
and other galaxies surrounding the wall, so that 
the aforementioned inflationary line element  is 
not  applicable. 

Another important gravitational effect is the mixing of 
the $\{zm\}$ components ($m=x,y$) of the graviton with the 
zero mode of the $\phi$ field on a  $(2+1)$-dimensional wall worldvolume,
which, in principle, leads to a ``mass" of the graviton
inside this layer. Let us show that this effect
is so small numerically that it can be safely neglected at   distances
of the size of the galactic disk or  larger.

The mixing term emerges  due to the interaction 
\beq
\int dz ~\sqrt{g}~g^{\mu\nu}~\partial_\mu
\phi~\partial_\nu \phi~.
\label{gphi}
\eeq
When one of the $\phi$'s is  substituted by the classical solution
(\ref {cl}), this term gives rise to an effective 
(2+1)-dimensional mixing of $\rho$ 
with the vector component of 
the metric (``graviphoton''), 
\beq
A^m(t,x,y)~\equiv ~\sqrt{m_d} ~\int_{-\Delta/2}^{\Delta/2} 
~dz ~g^{mz}(x, z)~M_P,\quad
m=0,1,2\,.
\label{A}
\eeq
This (2+1)-dimensional mixing takes the form 
\beq
m_d~{\eta\over M_P}~A^m(t,x,y)~\partial_m~\rho(t,x,y)~.
\label{mixing}
\eeq
The graviton propagation along the wall worldvolume is  
modified  due to this  mixing term.\footnote{In addition, there 
will  be induced terms for gravitons \cite {DGPind}
and gauge fields \cite {DGS} which could give rise 
to additional interactions \cite {DGPcc,DGPind,DGS}.
However, these effects are negligible in the present case
due to the low density of the galaxy and domain wall matter.}
This  leads to the constraint 
\beq
\eta~\ll  M_P~\left ({\Delta\over L}\right )~,
\label{eta}
\eeq 
where $L$ denotes the longitudinal size of the galactic luminous disk,
or any appropriate distance along the wall. 
Since the ratio $ {\Delta/ L}$ is typically of the order
$10^{-2} -10^{-3}$, the constraint (\ref {eta})
is not   stringent. We will see below that other 
considerations restrict the value of $\eta$ much more severely.
In fact, one can go to as large values of
$L$ as $L\sim 10^{27}$ cm
without violating the constraint (\ref{eta}).

The last issue to be discussed in this section is the back reaction of
matter on the wall.
Certainly, this interaction distorts the wall profile.
The  easiest way to picture the  distortion is to consider a 
single infinite domain wall on which a single finite-size 
galaxy disc is placed. 
Far away from the galaxy disc (along the wall) the effect of the 
distortion is negligible. As we come closer
to the galaxy the distortion could become noticeable.
The wall may swell but it cannot disappear;
 the trapping due to the $\chi$ component of the wall
is operative\footnote{The domain
wall profile  will also 
be distorted  with respect to that of the 
galaxy matter, because the domain walls and 
the matter density follow different laws of the cosmological expansion.
However, these effects are negligibly small in our model.}.

\section {Application to Flat Rotational Curves}

One of motivations  for the existence of dark matter 
is the experimental result on flat  rotation curves of
spiral galaxies \cite {velocity}. 
The measured rotational velocity $v$ can be related to the classical  
interaction potential  $V(r)$ as follows:
\beq
v^2~=~r~{dV(r)\over dr}\, .
\label{v}
\eeq
Experimental data \cite {velocity} show that
the velocity squared (after subtracting from it the part  due to 
the Newtonian attraction) 
is nearly constant 
at distances much larger than the radius of the galaxy disk,
\beq
v^2_{\rm exp}~\sim~{\rm constant}\, .
\label{const}
\eeq
The conventional  
Newton potential, $V(r)\sim M_G/r$, with the constant
 galaxy mass   $M_G$ does not satisfy (\ref {v})
provided that (\ref {const}) is valid. A standard way out is
to  assume that there is a halo of dark matter which 
surrounds  the galaxy disk, with the 
density distribution decreasing as $\sim 1/r^2$.
As a result,  the total mass $M_G$ is raising linearly with 
$r$ as long as we are inside the
dark matter halo, and (\ref {const}) is fulfilled. 

Dark matter can  have a few distinct components.
The density of luminous matter is estimated to be
$ \Omega_{\rm lum}~\sim~  \left ( 0.003 - 0.006 \right )\,.$
On the other hand,  Big Bang 
Nucleosynthesis (BBN) \cite {Wagoner} predicts the baryon density in the
ballpark (for discussions, see, e.g. \cite {Kolb}):
$\Omega_{\rm B}~\sim~\left (0.015-0.16\right )\,.$
Since 
$\Omega_{\rm lum} ~< ~\Omega_{\rm B}$,
there is a substantial amount of 
dark baryon matter in  the Universe.

It is believed that there is  
at least one additional very important component  of  dark matter, 
cold dark matter (CDM). 
CDM seems to be necessary for successful  structure formation
\cite {Structure}. Although, the CDM could be providing
dominant components in the elliptic galaxies, 
it is not excluded that the mechanism for flat rotational velocity 
curves of spiral galaxies is different. In particular, we explore
below the possibility that flat rotational curves in spiral galaxies
are due to the same domain walls which trap the galaxy matter
in the disk plane.

Let us now discuss dynamical aspects of our  mechanism in 
more detail.
The logarithmic correction to the 
potential  at galactic scales does  explain  the flat rotational curves,
\beq
\delta V(r)|_{\rm few~ kpc}~\simeq~C~{\ln}~r \, ,
\eeq
where $C\simeq v^2$. The logarithmic potential obviously
satisfies Eq. (\ref{v}) with the asymptotically
constant velocity.
In our galaxy $v\simeq 220~ {\rm km/s}$, hence
\beq
C~\simeq~5\times 10^{-7}\,.
\eeq
In general, $v\simeq (60$ to $300)~ {\rm km/s}$ and, thus, 
$C\sim (10^{-8}$ to $10^{-6})$. 

The authors of Ref. \cite {Milgrom} argue that 
the modified potential can arise   
due to a change of general relativity
(GR) at galactic scales. (A typical galactic size is of the order of
several units times
$10^{22}$ cm. We take its thickness outside the central domain, the bulge,
to be of the order of $10^{20}$ cm.)
Eventually this interesting proposal might  find some  theoretical
ground.  At present, we are aware of no possibility for
 constructing a theoretically
and phenomenologically  viable  modification 
of GR,   which would give rise to such a potential.
However, the logarithmic interactions can be obtained from domain walls,
as we argued above. Let us estimate the force due to this
logarithmic potential.
The new term generates an interaction
of  the field $\rho$ with  $\tilde\theta^\mu_\nu $,  an effective
``three-dimensional''  energy-momentum tensor
 measuring  surface energy density in the galactic 
disk, 
\beq
{\cal L}_{\rm int}^{\rho}~\simeq~ {1\over \sqrt{m_d}~M_* \Delta} 
~\rho~{\tilde \theta}^\mu_\mu~\equiv g~\rho~{\tilde \theta}^\mu_\mu~ ,
\label{rho}
\eeq
where 
\beq
\Delta ~\equiv ~{1\over m_{d}}~\sim~{1\over \sqrt{\lambda}~\eta}
~\sim~ 10^{20}~{\rm cm}\,,\qquad g=
\frac{\sqrt{m_{d}}}{M_*}\,,\qquad \tilde
\theta^\mu_\nu~\sim \theta^\mu_\nu~\Delta\,.
\eeq
Here, $\Delta$ is the wall thickness, of 
the order the transverse width of the 
luminous disk. Let us make a comment here. When we go to  
the distances which are larger than the disk diameter 
the interactions along the disk plane are mediated 
by the exchange of a zero-mode field $\rho$. 
All other $(2+1)$-dimensional massive  modes of the 
$\phi$ field are decoupled at those 
distances. However, at distances which are smaller than the disk diameter,
at the solar system distances for instance,
all those massive states give rise to sizable contributions. 
The net result of these states is equivalent to 
the exchange of a single $(3+1)$ dimensional scalar mode $\phi$. 
This exchange is suppressed by $1/M_*$ and its effect is negligible
for the solar system data.

We come to the following scenario.
At the scale of  distances  less than the galactic thickness
(i.e. $\ll 10^{20}$ cm), in particular, in the Solar system,
the impact of the field $\phi$ reduces to a small ($10^{-3}$ to 
$10^{-4}$, see
below) correction to the Newtonian law. This correction shows up in the
form of discrepancy between the measurements for the static and
relativistic matter (e.g. light deflection). At such distances the mass of
the $\phi $ quantum can be neglected, and it acts as a massless
3+1-dimensional field. 
This field couples to the galactic matter with yet-to-be-determined 
coupling $1/M_*$. 

However, at distances $\gg \Delta$, in particular at distances
of the order
the  galactic disk size or larger, one  can  neglect the thickness of the
galactic disk/domain wall.  Then, $\rho$  behaves as a  2+1-dimensional
excitation that mediates 
the logarithmic potential,
\beq
\delta V(r)|_{\rm few ~~kpc}~\simeq~M_G~g^2~{\rm ln}~r~.
\eeq
Using the known value of  the rotational velocity we find 
\beq
M_G~g^2~\sim~C~\sim~5\times 10^{-7}.
\eeq
This can be used to 
determine $M_*$. Using the mass of the luminous galaxy for $M_G$
one would underestimate $M_*$. Instead,  
we   take into account 
dark baryon matter in the galaxy disk. This gives (see Eq. (11))
\beq
M^2_*~\sim~\left (10^{41}- 10^{42}\right )~{\rm GeV}~\simeq~
\left (10^3-10^4  \right )~M^2_P~.
\eeq
Therefore,  we conclude that the interaction of the 
four-dimensional scalar with matter is sufficiently suppressed.  
It does not contradict the Solar system data 
such as bending of light rays and
precession of the Mercury perihelion which now are measured
to one percent accuracy.
 
The modification of the Newton law discussed above will lead to a
peculiar ``two-component'' interaction between distant galaxies located
inside distinct wall cells (Fig. 1). First, there is a conventional
Newtonian attraction through the bulk (due to conventional
four-dimensional gravitons). This is not the end of the story, however.
In addition, our dilaton gravity will propagate through the domain wall
network\footnote{We are grateful to Tonnis ter Veldhuis who posed this
question.}. Consider, for instance, two galaxies on adjacent domain walls
connected by a common junction. Assume the distance of each galaxy to the
junction is $R$. Then each of the galaxies will experience an additional
(non-Newtonian) attraction force towards the junction of order of
$(M_G/M_*)^2 (R\Delta)^{-1}$. This force falls off with the distance as
the first rather than the second power of $R$. It becomes comparable
with the Newtonian force at distances $R\sim 10^4 \Delta \sim
10^{24}$cm. 
If the number of the walls $N$ joined in the given junction
is large, say,  $N\sim 10$, the extra  force will
   be further suppressed by $1/N$ because the force line flux
approaching the junction from one wall, after the junction
 will be shared by $N-1$ walls. This would shift the critical $R$
even to larger distances, $R\sim 10^{25}$cm.
 
Typically distant galaxies will be separated by several (perhaps, many)
junctions. If the number of junctions is $k$, the additional suppression
factor one gets is $(1/N)^k$.
 
The ``simulated" attraction of the galaxies to the junction will tend to
distort the form of the junctions themselves. This effect can be easily
estimated by comparing the energy (per unit area) added to the domain
wall due to the dilaton gravity induced by the given galaxy,
$$
\sigma_{DG} ~\sim ~\left(\frac{M_G}{M_*}\frac{1}{\Delta  R}
\right)^2~\Delta ~\lsim 10^{-11}~ {\rm GeV}^3~,
$$
if $R\gsim 10^3 \Delta \sim 10^{23}$ cm. Since in our model the wall
tension
$\sigma_{DW} \lsim 10^{-11} ~{\rm GeV}^3$, the distortion is not drastic
and can be neglected at the level of accuracy we maintain here.   

The next comment concerns the Tully-Fisher (TF) 
relation \cite {TullyFisher}.
This is a connection  between the flat rotational velocity  and
luminosity  $L$ of a galaxy:
$L~\propto~v^{\alpha}$. The coefficient $\alpha$ 
is fit by the data and ranges for various galaxies 
in the interval $\alpha=(2.5-4)$.
If the luminosity were a linear function of 
the galaxy mass $M_G$, the TF relation would give rise to
the dependence  $v^{\alpha}~\propto~M_G$. This would certainly 
exclude our scenario, since the latter predicts $v^2~\propto~M_G$.
However, the relation between the luminosity and the mass 
does not need to be linear. The total mass of the galaxy (without the 
CDM contribution) is composed of the luminous disk matter mass $M_d$, 
the mass of the gaseous
matter $M_g$, the mass of the central bulge region of the 
galaxy $M_b$ and, finally, of the dark baryon mass. 
Although  the luminous disk mass could be 
linearly dependent on  the luminosity, the other three components
do not have to. Moreover, these components 
constitute a very important part of the galaxy mass. 
Just for the demonstrational purposes we present the mass contents of a few 
spiral galaxies (for detailed discussions, see, e.g. \cite {List}): 
$(UGC2885, ~M_g=5.0~, ~M_d=25.1~, ~M_b=5.7),~     
(NGC5533, ~M_g=3.0~, ~M_d=2.0~, ~M_b=17.0),~
(NGC6674, ~M_g=3.9~, ~M_d=2.5~, ~M_b=15.5 ),
(NGC5907, ~M_g=1.1~, ~M_d=7.2~, ~M_b=2.5 )$. All masses here are given in the 
units of $10^{10}~M_{\rm Sun}$. Therefore, it is conceivable 
to assume that  the relation  $v^2\propto~M_G$, predicted by 
the present scenario,  does not contradict the known data.

So far  we  assumed that
only a single  galaxy disk can be embedded in a  given domain wall.
However, this condition is not necessary. 
For instance, one could imagine 
a pair of  nearby galaxies which are  
rotating with respect to each other  
inside one and  the same domain wall. 
If so, there is an additional logarithmic force
between them due to the exchange of $\rho$. 
Due to this logarithmic force,
the nearby binary galaxies would  seem to be  
``confined'' to each other. 
This can be used to explain the old puzzling data 
of Ref. \cite {Einasto} where  it was observed that 
binary galaxies
attract  each other with forces stronger  than those exerted by 
their disks. 
On the other hand, distant galaxies can and should  be located inside  
distinct domain walls. Then, the  logarithmic potential between them 
will not be  present and they will be interacting by means 
of Newtonian force.

\section{Other data}

Let us now check whether this  type of domain wall scenarios   
are allowed by cosmological and astrophysical data. 

\begin{itemize}

\item{Domain Wall Domination}

First we have to make sure that the tension of the domain wall
is small enough so that its contribution to the 
density of the Universe is not large and the evolution of the Universe
is not domain wall dominated. The standard estimate 
\cite {ZKO} leads:
\beq
\Omega_{\rm DW}~\sim~{\s\over 10^{-5}~{\rm GeV}^3}~.
\eeq
Given that $m_d\sim \left (10^{-33} -10^{-32}\right )~{\rm GeV}~$
and assuming that $\Omega_{\rm DW}\ll 1$, we get
$\eta \ll 10^{13}~{\rm GeV}$.

\item{The CMBR anisotropy}

The presence of the domain wall network in the Universe would lead to 
an
anisotropy in Cosmic Microwave Background Radiation. 
The CMBR temperature anisotropy ${\delta T/ T}$ is 
measured with very high accuracy. Therefore, the anisotropy
introduced by the domain walls should be constrained as follows:
\beq
{\delta T\over T}~\sim~ {\s\over H_0 ~M_P^2}~<~10^{-4}~.
\eeq
This gives rise to a more stringent bound on $\eta$,
$$
\eta < 10^{11}~{\rm GeV}\,.$$
 Note that if 
$\eta$ is close to its upper bound, then the coupling constant
$\lambda$ is extremely small, $\lambda\sim 10^{-88}$.
Therefore, self-interactions of the $\phi$ field 
are harmless.  

There are additional constraints which should be imposed on 
the mass and decay constant of $\phi$. These come from: 

\item{Star cooling}

\item{Big Bang Nucleosynthesis}

In the former case one must make sure that 
the stars do not cool too fast due to the emission of 
energy in the form of the light $\phi$ quanta.
Moreover, for the purpose of successful BBN the decay
$\phi\rightarrow \gamma\gamma$ should be suppressed enough 
to avoid overproduction
of entropy which would spoil the standard BBN scenario.
Using the analysis of \cite {Voloshin}
one can see that the scalar field with the mass  
$m_{d}\sim 10^{-32}$ to $10^{-33}\,\,\, {\rm GeV}~$
and the constant $M_*\sim~10^2~M_P$
 satisfies these constraints.

\end{itemize} 

\vspace{0.1in}

\section{Cosmological evolution}

A  few words are
in order regarding the cosmology of the ``dilatonic"  
wall formation.  The domain walls that can 
survive till the present epoch must have been formed
 after the inflation.  This 
implies that the discrete symmetry must be restored either during 
the inflation or, at least, during reheating. For the ``dilatonic" type walls 
under consideration this issue is somewhat subtle, as we will now 
discuss.  

First, during the inflation the VEV of $\phi$ cannot vanish, due to 
the fact that expectation value of $\theta^{\mu}_{\mu}$ is nonzero and
it 
generates a tadpole for the $\phi$ field. As a result,
the  pre-existing domain walls 
would be inflated away. If the only coupling to matter is through the term
$
  e^{\phi/M_{*}}~ \theta_{\mu}^{\mu},
$
the symmetry cannot be restored after reheating either, since the thermal 
average of the $\theta^{\mu}_{\mu}$ vanishes and $\phi$ is never in the 
thermal equilibrium. Thus, we are left with no walls. 

A way out would be 
to add some additional, ${\bf Z}_N$-preserving interactions 
which would couple   $\phi$ 
to  
matter fields that are in thermal equilibrium during reheating. Let 
$X$
be such  scalar  matter field. Then the relevant interaction is
\begin{equation}
{\phi^2\over M_P^2} X^4 \,.
\label{phichi}
\end{equation}
We choose the sign of this coupling to be positive.
Although $\phi$ is not in the equilibrium, the above coupling generates an 
effective potential for it, which can restore the symmetry. The relevant 
diagrams are ``butterfly" diagrams with two $X$ loops and external
$\phi$ legs \cite{dkbs}. Each $X$ loop contributes a  $T^2$ factor 
($T$ being the temperature), so that the 
resulting effective operator in the free energy has the form
\begin{equation}
 \phi^2 {T^4 \over M_P^2}\,,
\end{equation}
with the positive coefficient.
Since in the radiation dominated epoch ${T^4\over M_P^2} \sim H^2$,
the field  $\phi$ 
gets a positive mass term of the order of the Hubble parameter.
At the same time, there is no tadpole since  $\theta_{\mu}^{\mu}$ 
vanishes. Thus, the discrete symmetry may be restored until the zero 
temperature potential starts dominating again, and the phase transition 
takes place with the subsequent wall formation. 

Note that the coupling 
(\ref{phichi}), if included in a non-universal way, can induce an 
unacceptably large zero-temperature mass renormalization of  $\phi$, 
which should be readjusted to zero along with the possible quantum gravity
corrections. This does not add  additional fine-tuning to the model.

\section{Comments on the literature}

It should be noted that
models with ultralight 
(pseudo)Goldstone bosons developing domain walls
and characterized by phase transitions at late time, after the 
decoupling of the microwave
background, had been  suggested long ago \cite{Hill}.
The Compton wave length of the ultralight boson
considered, was in the ballpark of
1 to 100 Mpc, its mass being protected by a 
continuous U(1) symmetry, of which the  ultralight boson
in question is the Goldstone boson. It was found that
this set-up may be helpful from the phenomenological standpoint,
for the large-scale structure formation, a welcome feature. 

While this model shares certain common elements with ours,
differences are crucial.
The main distinction
is that the coupling of the 
ultralight boson \cite{Hill} with fermions is nonuniversal and 
pseudoscalar (which, naturally, does
not allow for coherent effects inherent to the
 dilaton universally coupled
to the trace of the energy-momentum tensor).\footnote{$CP$ violation 
might turn the pseudoscalar coupling into scalar \cite{Hill2}.
The amount of $CP$ violation is severely limited, however, by the
neutron dipole moment data.} 
It remains to be seen whether 
our dilaton-based model enjoys  the same favorable 
environment for the large-scale structure formation as that of Ref. 
\cite{Hill}.

An {\em ad hoc}  logarithmic long-range potential
coupled to baryon number, as a possible explanation of
the constancy of the rotational curves, was also discussed in
\cite{KB}. Since this potential was assumed to exist 
in the three-dimensional bulk, this led to irreconcilable 
contradictions with the microlensing data.
The authors themselves noted that their construction was ruled out.
None of the drawbacks of \cite{KB} exist in our scheme.

\section{Discussion}

We presented a scenario in which spiral galaxies
have their natural habitat inside very fat domain walls.
The existence of these walls gives rise to trapping of  
 matter within the wall 
which may provide a natural explanation for the
very
formation of  flat spiral galaxies. 
In addition, a scalar particle living on the walls (zero modes),
generates a logarithmic potential which takes over the Newtonian $1/r$
at distances $\gsim 10^{22}$ cm. This explains the constancy of the tails of
the rotational curves in  the spiral galaxies.

Although our mechanism ensuring a logarithmic
component in the attraction potential of matter is very simple, surprisingly
it seemed to escape attention of the previous investigators.
As far as we can see, it contradicts no existing data.
However, there
are several issues which should be studied further.
Among those are the questions: 
Are there subtler effects which might limit
the applicability of this mechanism?  Do ``exceptional'' galaxies which are
perpendicular to the walls exist? 
Is there a place in this picture for elliptic galaxies and
galaxy  clusters? (Near the junctions?)
Could the velocity dispersion in   the elliptic galaxies be 
explained by the presence of dark baryons alone? 
All these questions should be analyzed further.
However, since these 
issues cannot be studied analytically, but rather require 
involved numerical simulations, we did  not address them here. 

To our mind, a hint that there might be some truth in the
suggested picture is
 the remarkable numerical proximity of the
parameter $M_*$ obtained in Sec. 4 from the galactic parameters, to the
Planck scale. It is unlikely that it is accidental.
Another positive feature that makes us optimistic is the fact that
we found a set-up ensuring  ``almost no  renormalization"
environment for the ultralight scalar
(at least, as far as the SM loops are concerned). The universal coupling
to the trace of the energy-momentum tensor, in conjunction with the scale
Ward identities, provides a natural protection.
The parameters of the scale symmetry breaking in the dilaton sector we
need for the success of the model are such that this protection is operative.

We would like to stress that 
the mechanism outlined  in 
the present paper can be confronted with experiment  
in the near future.
Indeed, the coupling of the $\phi$ field to  matter
is suppressed with respect to the gravitational coupling
only by a three to four orders of magnitude. This is a borderline 
where the equivalence principle, as well as 
relativistic gravitational effects, are tested
experimentally. Further improvements in the measurements 
by one or two orders of magnitude could  be decisive 
for this proposal.
 
There is another tantalizing experimental possibility \cite{Nuss}.
Assume, in a given spiral galaxy two independent measurements
are performed: gravitational lensing and the rotational curve.
One could reconstruct the dark matter distribution that would fit the
result of the gravitational lensing measurement. Assume the
reconstruction has sufficient precision that would enable one to say
that the amount of dark matter obtained from gravitational lensing
is insufficient to describe the rotational curve. This would be a very
strong argument in favor of our conjecture.

\vspace{0.3in}

{\bf Acknowledgments}

\vspace{0.1in}

The authors are grateful to Z. Berezhiani, 
G. Farrar, Z. Kakushadze, 
N. Kaloper, R. Konoplich, S. Nussinov, K. Olive, M. Porrati, 
M. Pospelov, G. Senjanovi\'c, E. Skillman, 
T. ter Veldhuis, A. Vilenkin, M. Voloshin, 
L. Williams, and  M. Zaldarriaga for useful discussions.
We acknowledge communications with C. Hill, W. Kinney and A. Kosowsky
who pointed out to us several relevant
references which were, unfortunately, omitted in the first version.

The work of G.D. was supported in part by David and Lucille 
Packard Foundation 
Fellowship for Science and Engineering, 
and by Alfred P. Sloan Foundation 
Fellowship and by NSF grant PHY-0070787. Work of G.G. and M.S. was 
supported by DOE grant DE-FG02-94ER408.

\end{document}